\documentclass[aps,pra,twocolumn,a4paper,floatfix,superscriptaddress, nofootinbib]{revtex4-2}
  \usepackage[T1]{fontenc}
  \usepackage{amsmath}
  \usepackage{amsthm}
  \usepackage{amsbsy}
  \usepackage{amssymb}
  \usepackage{enumitem}
\usepackage{lipsum}
\usepackage[colorlinks=true,urlcolor=blue,citecolor=blue,linkcolor=blue]{hyperref}
\usepackage{graphicx, color}
\usepackage[dvipsnames,table,xcdraw]{xcolor}
\usepackage{braket}
\usepackage{orcidlink}
\usepackage[caption=false]{subfig} 
\usepackage{physics}

\DeclareMathOperator{\BesselJ}{J}

\begin{document}
\title{Quantum-limited estimation of atmospheric turbulence via spatial mode decomposition}

\author{A. Hrebeniuk \orcidlink{0009-0003-0224-1625}}
\affiliation{Quantum Optics and Quantum Information Group, Bogolyubov Institute for Theoretical Physics of the National Academy of Sciences of Ukraine, Vulytsia Metrolohichna 14b, 03143 Kyiv, Ukraine}

\author{M. Klen \orcidlink{0009-0009-3030-6450}}
\affiliation{Quantum Optics and Quantum Information Group, Bogolyubov Institute for Theoretical Physics of the National Academy of Sciences of Ukraine, Vulytsia Metrolohichna 14b, 03143 Kyiv, Ukraine}

\author{I. Karuseichyk \orcidlink{0000-0003-4981-6230}}
\affiliation{Laboratoire d’Optique Appliqu\'e (LOA), CNRS, \'Ecole polytechnique, ENSTA, Institut Polytechnique de Paris, Palaiseau, France}

\author{N. Treps \orcidlink{0000-0002-1413-3715}}
\affiliation{Laboratoire Kastler Brossel, Sorbonne Universit\'e, ENS-Universit\'e PSL, CNRS, Coll\`ege de France, 4 Place Jussieu, F-75252 Paris, France}

\author{A. A. Semenov \orcidlink{0000-0001-5104-6445}}
\affiliation{Quantum Optics and Quantum Information Group, Bogolyubov Institute for Theoretical Physics of the National Academy of Sciences of Ukraine, Vulytsia Metrolohichna 14b, 03143 Kyiv, Ukraine}
\affiliation{Department of Theoretical and Mathematical Physics, Kyiv Academic University, Boulevard Vernadskogo  36, 03142  Kyiv, Ukraine}
\affiliation{Department of Mathematics, Kyiv School of Economics, Vulytsia Mykoly Shpaka 3, 03113  Kyiv, Ukraine}

\begin{abstract}
We establish the ultimate precision limit for estimating
%atmospheric scintillometry by formulating an estimation theory for 
the optical spatial coherence radius (Fried parameter) within a quantum metrological framework. 
In the weak field regime, we show that spatial-mode decomposition---originally introduced for superresolution imaging---enables substantially more precise estimation than conventional direct imaging when the receiver aperture is smaller than the coherence radius.  
\end{abstract}

\maketitle

\section{Introduction}
\label{Sec:Intro}

When light propagates through a turbulent atmosphere, it is affected by random fluctuations in the refractive index \cite{Tatarskii2016,tatarskii1971,Fante1975,Fante1980,Andrews_book}.
Turbulent effects pose challenges for both classical and quantum optical applications, leading, for example, to fluctuating signal losses in free-space communication \cite{semenov09,vasylyev12,vasylyev16,vasylyev18,klen2023,semenov2025} and degradation of optical resolution \cite{Fried1966,roddier1981,Kopeika1998,Haik2007}.
A specific parameter of interest for optical turbulence estimation is the spatial coherence radius $r_0$, which is closely related to the Fried parameter~\cite{Fried1966}.

A straightforward approach to estimating $r_0$ is direct imaging (DI), which infers it from the light-spot size of a point-like source; see, e.g., Ref.~\cite{roddier1981} for a review.
This technique, however, is limited by diffraction when the receiver aperture is significantly smaller than the coherence radius $r_0$.
Modern scintillometric techniques \cite{Lombardi2014,li2022} include differential image motion monitoring \cite{sarazin1990,tokovinin2002dimm,berdja2010,aristidi2014}, wavefront sensing \cite{dayton1992, wilson2002, butterley2006, silbaugh1996, Andrade2018, zuraski2020}, and methods based on irradiance fluctuations \cite{wang1978, foken2021, Charnotskii2021}, along with other approaches \cite{vonderLuhe1984, tokovinin2010, avila1997,Ren2022,tokovinin2002}.

A conceptual analysis of methods for estimating $r_0$ enables their interpretation within the framework of parameter-estimation theory.
Electromagnetic radiation, stochastically perturbed by atmospheric turbulence, is analyzed by measuring an observable whose probability distribution depends on the turbulence parameters.
For single-photon detection, this observable may be the photon image-plane position $\mathbf{r}=\begin{pmatrix} x & y \end{pmatrix}$, as in the DI scenario.
The task is thus to estimate the parameter $r_0$ based on a sample $\{\mathbf{r}_i\}_{i = 1}^N$ and a statistical model given by the probability distribution $P(\mathbf{r} | r_0)$, where $N$ is the sample size.
This constitutes a typical parameter-estimation problem \cite{Kay93,lehmann2006,voinov2012}.
For (asymptotically) unbiased estimators, their precision (variance) is bounded by the Cram\'{e}r--Rao  (CR) inequality \cite{rao45,cramer_book}, 
    \begin{align}
        \Delta r_0^2\geq\frac{1}{N\mathcal{I}_{\mathbf{r}}(r_0)},
    \end{align}
where $\mathcal{I}_{\mathbf{r}}(r_0)$ is the Fisher Information (FI) \cite{fisher1925} associated with photon-position measurements.

More broadly, estimation of atmospheric turbulence can be considered within the powerful framework of quantum metrology~\cite{Helstrom1969,Helstrom_book,Braunstein1994,Giovannetti2006,PARIS2009,Giovannetti2011,Pezze2018}. 
In this approach, the density operator of the electromagnetic radiation, $\hat{\rho}$, depends explicitly on $r_0$. 
Maximizing FI over all measurements yields the ultimate measurement-independent quantum CR bound associated with the Quantum Fisher Information (QFI) $\mathcal{K}(r_0)$, such that $\mathcal{I}(r_0)\leq\mathcal{K}(r_0)$.

While a quantum-metrological approach based on specially prepared quantum states was considered in Ref.~\cite{Yu2024}, we instead employ simple passive light sources.
Our idea is to tailor the spatial mode decomposition (SpaDe) method~\cite{Tsang2016,Nair2016,Tsang2017,Rehacek2017,Boucher2020,Rouviere24} to estimate $r_0$ using radiation from a point-like source in the far-field regime. 
The SpaDe approach is based on a linear decomposition of the incident light into orthogonal spatially structured modes, followed by separate detection of each mode.
This technique allows one to overcome the Rayleigh diffraction limit in imaging. 
In analogy with that problem, we calculate the QFI for the spatial coherence radius $r_0$ and compare it with the FI for DI and SpaDe detection. 
Estimating $r_0$ becomes particularly challenging if it significantly exceeds the receiver aperture diameter. 
In this regime, efficient extraction of the available information is therefore especially crucial. 
We show that SpaDe detection approaches the quantum limit for estimating weak turbulence, while DI discards a significant fraction of the information. 
This points to a practical route for precise estimation of atmospheric parameters with finite-aperture receivers. 

\section{Quantum state of light}
\label{Sec:QuantumState}

Let us consider radiation from a passive, monochromatic point-like thermal source \cite{zmuidzinas2003,labeyrie2006,goodman2015,mandel1959,mandel1995,gottesman2012,tsang2011} that may be located either within the atmosphere or far beyond it.
In the far-field regime, the spherical wave after propagation through the atmosphere can be approximated by a turbulence-perturbed plane wave, which we associate with a spatial mode (field amplitude) $u^{\textrm{(ap)}}(\mathbf{r})$ confined to and normalized over a finite area $S$. 
The light then passes through a circular aperture of area $S_{\mathrm{ap}} < S$, is focused by a thin lens onto the image plane, and is analyzed using either the DI or SpaDe technique.

The unnormalized field amplitude $u^{\textrm{(im)}}(\mathbf{r})$ in the image plane is related to $u^{\textrm{(ap)}}(\mathbf{r})$ via the Huygens-Fresnel diffraction integral \cite{born2013,goodman2005},
		\begin{align} \label{Eq:ImageField}
			u^{\textrm{(im)}}(\mathbf{r}) =
			\frac{e^{i 2\pi f / \lambda}}{i \lambda f}
			\int_{\mathcal{A}} \dd^2\mathbf{r}^\prime \,
			u^{\textrm{(ap)}}(\mathbf{r}^\prime) \,e^{- \frac{2\pi i}{\lambda f} \mathbf{\mathbf{r}^\prime} \cdot \mathbf{r}}.
		\end{align}
Here $\lambda$ and $f$ are the wavelength and focal length of the lens.
The integration is performed over the finite aperture opening $\mathcal{A}$. 
The field norm,
		\begin{align}\label{Eq:eta}
			\eta=\int_{\mathbb{R}^2}\dd^2\mathbf{r} |u^{\textrm{(im)}}(\mathbf{r})|^2\in[0,1]
		\end{align}
can be interpreted as the aperture transmittance.

Let $n_{\mathrm{th}}$ be the mean photon-number arriving to the aperture plane.
In a weak-field regime ($\eta n_{\mathrm{th}}\ll 1$), the instantaneous quantum state in the image plane can be approximated as a statistical mixture of the vacuum and single-photon states,
    \begin{align}\label{Eq:DensOp_Inst}
        \hat{\rho}_{\mathrm{inst}}&=(1-\eta n_{\mathrm{th}})\dyad{\mathrm{vac}}{\mathrm{vac}}\\
       &+n_{\mathrm{th}}\int_{\mathbb{R}^4}\dd^2\mathbf{r}_1\dd^2\mathbf{r}_2\,
       u^{\textrm{(im)}}(\mathbf{r}_1)\, [u^{\textrm{(im)}}(\mathbf{r}_2)]^{\ast}\,\dyad{\mathbf{r}_1}{\mathbf{r}_2}.\nonumber
    \end{align}
Here $\ket{\mathbf{r}}$ denotes the single-photon state localized at the transversal point $\mathbf{r}$ of the image plane.

Atmospheric turbulence randomly perturbs the field amplitude, rendering $\eta$ a random variable.
Averaging the instantaneous density operator in Eq.~(\ref{Eq:DensOp_Inst}) over atmospheric realizations (denoted by $\left\langle\ldots\right\rangle$) yields the image-plane density operator,
    \begin{align} \label{Eq:State}
        \hat{\rho}=\left(1-\left\langle\eta\right\rangle n_{\mathrm{th}}\right)\dyad{\mathrm{vac}}{\mathrm{vac}}+\left\langle\eta\right\rangle n_{\mathrm{th}}\,\hat{\rho}_1,
    \end{align}
where 
    \begin{align} \label{Eq:DensOp_1}
    	\hat{\rho}_1 = \int_{\mathbb{R}^4} \dd^2 \mathbf{r}_1 \dd^2 \mathbf{r}_2\Gamma_2^{\textrm{(im)}} (\mathbf{r}_1, \mathbf{r}_2) \dyad{\mathbf{r}_2}{\mathbf{r}_1}
    \end{align}
is the single-photon density operator, expressed in terms of the second-order field correlation function in the image plane,
    \begin{align}\label{Eq:Gamma2_im_gen}
       \Gamma_2^{\textrm{(im)}} (\mathbf{r}_1, \mathbf{r}_2)=\left\langle\eta\right\rangle^{-1}\left\langle\,u^{\textrm{(im)}}(\mathbf{r}_1)  [u^{\textrm{(im)}}(\mathbf{r}_2)]^{\ast}\right\rangle. 
    \end{align}
Here $\langle\eta\rangle$ plays the role of the normalization factor, required for a consistent definition of the density operator~$\hat{\rho}_1$.

For a source located either far beyond or within the atmosphere the second-order field correlation function in the aperture plane, within the Kolmogorov--Obukhov turbulence model, is given by (see Refs.~\cite{Tatarskii2016,tatarskii1971,Tatarskii1980,Andrews_book})
    \begin{align} \label{Eq:Gamma2_ap}
        \Gamma_2^{\textrm{(ap)}}(\mathbf{r}_1, \mathbf{r}_2) = \frac{1}{S} \exp\left[-\left(\frac{|\mathbf{r}_1-\mathbf{r}_2|}{r_0}\right)^{5/3}\right].
    \end{align}
Here $r_0$ is the spatial coherence radius, which is the parameter to be estimated.
The model given by Eq.~(\ref{Eq:Gamma2_ap}) suffices to capture the core physics, while extensions are discussed in Appendix~\ref{App:A}. 

Combining Eqs.~(\ref{Eq:ImageField}), (\ref{Eq:Gamma2_im_gen}),  and (\ref{Eq:Gamma2_ap}), we get
     \begin{align}\label{Eq:Gamma2_im}
		&\Gamma_{2}^{(\mathrm{im})} (\mathbf{r}_1, \mathbf{r}_2)
		=\frac{1}{\lambda^2 f^2 S_{\mathrm{ap}}} \int_{\mathcal{A}}\mathbf{d}^2 \mathbf{r}^{\prime} \int_{\mathcal{A}} \mathbf{d}^2 \mathbf{r}^{\prime\prime} 
		\\
		&\times
		\exp\!\left[
		- \frac{2 \pi i}{\lambda f} \left(\mathbf{r}^{\prime} \cdot \mathbf{r}_1 -\mathbf{r}^{\prime\prime} \cdot \mathbf{r}_2\right) 
		\right]          
		\exp\!\left[ -\left(\frac{\left|\mathbf{r}^{\prime} - \mathbf{r}^{\prime\prime} \right|}{r_0}\right)^{5 / 3} \right],\nonumber
	\end{align}
and $\langle\eta\rangle = S_{\mathrm{ap}}/S$.
Therefore, in Eq.~(\ref{Eq:State}) only the single-photon density operator $\hat{\rho}_1$ depends on the parameter $r_0$.
Hence, postselecting single-photon events entails no loss of information about $r_0$.

\section{Gaussian approximation}
\label{Sec:Gaussian}

The analysis of SpaDe superresolution imaging \cite{Tsang2016, jose2021, Giacomo2021} typically employs a model based on the soft-aperture approximation \cite{zhang2007}. 
In the context of estimating atmospheric turbulence, this approximation simplifies the second-order field correlation function $\Gamma_2^{\textrm{(im)}} (\mathbf{r}_1, \mathbf{r}_2)$ [see Eq.~(\ref{Eq:Gamma2_im})] by replacing it with a Gaussian function $\Gamma_2^{\textrm{(G)}}(\mathbf{r}_1, \mathbf{r}_2)$, 
    \begin{align}\label{Eq:G}
      \Gamma_2^{\textrm{(G)}}(\mathbf{r}_1, \mathbf{r}_2) =
      \frac{1}{2 \pi \theta^2}
      \exp\!\left[
        -\frac{1}{2}
        \begin{pmatrix}
          \mathbf{r}_1^{T} & \mathbf{r}_2^{T}
        \end{pmatrix}
        \Sigma
        \begin{pmatrix}
          \mathbf{r}_1 \\[2pt] \mathbf{r}_2
        \end{pmatrix}
      \right].
      \end{align}
Here the matrix $\Sigma$ is defined as
\begin{align}
  \Sigma =
  \frac{1}{4 \theta^2 \sigma^2}
  \begin{pmatrix}
    \theta^2 + \sigma^2 & \sigma^2 - \theta^2 \\ 
    \sigma^2 - \theta^2 & \theta^2 + \sigma^2
  \end{pmatrix}
  \otimes
  \mathbb{I}_2,
  \label{eq:A}
\end{align}
with $\mathbb{I}_2$ being the $2\times2$ identity matrix.  

This model explicitly depends on two parameters, $\theta^2$ and $\sigma^2$, directly related to the variances of the transverse photon position and transverse wave vector $\begin{pmatrix} k_x & k_y \end{pmatrix}$, respectively: $\theta^2=\langle \Delta x^2\rangle=\langle \Delta y^2\rangle$ and $\sigma^2=(4\langle \Delta k_x^2\rangle)^{-1}=(4\langle \Delta k_y^2\rangle)^{-1}$.
The parameter $\sigma$ corresponds to the characteristic size of the point-spread function (PSF), i.e., the light spot in the absence of turbulence, whereas $\theta$ characterizes the light-spot size averaged over atmospheric realizations.
These parameters are constrained by the uncertainty relation $\theta^2 \ge \sigma^2$, and the equality $\theta^2 = \sigma^2$ is attained only in the absence of atmospheric turbulence, when the state becomes pure. 

It is natural to define the parameters $\theta^2$ and $\sigma^2$ as those that maximize the quantum fidelity $\mathcal{F}\left(\hat{\rho}_1, \hat{\rho}_G\right)$ \cite{jozsa1994,nielsen2010} between the exact state $\hat{\rho}_1$ and its Gaussian approximation $\hat{\rho}_G$.
In contrast, Refs.~\cite{zhang2007,Tsang2016} determine the Gaussian parameters by fitting the intensity distribution, i.e. $\Gamma_2^{(\mathrm{im})}(\mathbf{r},\mathbf{r})$ and $\Gamma_2^{\textrm{(G)}}(\mathbf{r}, \mathbf{r})$. 
The results of the numerical evaluation of the quantum fidelity and its optimization (see Appendix~\ref{App:B}) allow us to find the parameters $\theta^2$ and $\sigma^2$ as
 \begin{align}\label{Eq:curve1}
   \theta^2
  \approx \left( 2.5 + \frac{1.2D^2}{r_{0}^{2}} \right)
   \frac{\lambda^{2} f^{2}}{2 \pi^{2} D^{2}},\quad \sigma^2\approx\frac{2.5 \lambda^{2} f^{2}}{2 \pi^{2} D^{2}},
 \end{align}
where $D$ is the aperture diameter. 
Remarkably, only the parameter $\theta^2$ depends on the turbulence strength and therefore constitutes the parameter to be estimated from the measurement data.

\section{Fisher information for soft-aperture model}
\label{Sec:FisherInfSoft}

\begin{figure}[ht!]
    \centering
    \includegraphics[width=\linewidth]{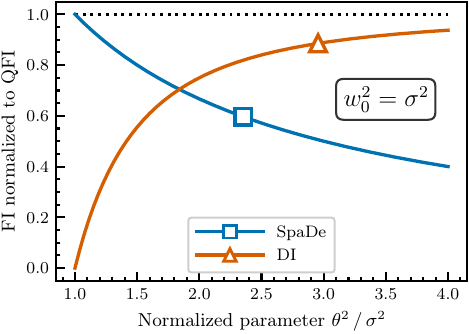}
    \caption{\label{Fig:Gaussian_theta} Ratio of the FI to QFI, $\mathcal{I}\left(\theta^2\right)/\mathcal{K}\left(\theta^2\right)$, as a function of the parameter $\theta^2$, for DI and SpaDe. }
\end{figure}

The Gaussian model is particularly useful because it allows one to derive explicit analytical expressions for the FI and the QFI, facilitating further analysis.
Adapting the results of Refs.~\cite{hubner1992,Braunstein1994,braunstein1996,marian2016,pinel2013}, we obtain the QFI for the parameter $\theta^2$ as
    \begin{align}\label{Eq:QFI_gauss}
        \mathcal{K}\left(\theta^2\right)= \frac{1}{\theta^{4} - \sigma^{4}}.
    \end{align}
Clearly, the QFI diverges in the absence of atmospheric turbulence, where $\theta^2=\sigma^2$, and decreases as the scintillation strength, i.e. the parameter $\theta^2$, increases.
The FI for DI is given by 
    \begin{align}\label{Eq:FI_Direct_G}
        \mathcal{I}_{\mathbf{r}}\left(\theta^2\right) = \frac{1}{\theta^{4}}.
    \end{align}
It remains finite in the turbulence-free limit and approaches the QFI as $\theta^2$ increases, i.e., as the scintillations become stronger; see Fig.~\ref{Fig:Gaussian_theta}.
This implies that DI is ineffective in the weak-scintillation regime.
For details of calculations, see Appendix~\ref{App:C}.

Next, we consider the SpaDe scheme, assuming that the mode sorter decomposes the optical field into two modes: the Gaussian with beam waist radius $w_0$ and its orthogonal complement---a configuration commonly referred to as binary SpaDe.
The FI for this scenario is given by
    \begin{align}\label{Eq:SpaDe_Gauss}
        \mathcal{I}_{\mathrm{SpaDe}}\left(\theta^2\right) = \frac{4 \, \sigma^2 \, w_0^2}{(\theta^2 + w_0^2)^2 \, [\theta^2 \sigma^2 + (\theta^2 - 3 \sigma^2) \, w_0^2 + w_0^4]}.
    \end{align}
We are particularly interested in the regime of the weak turbulence, i.e. $\theta^2=\sigma^2(1+\varepsilon)$, where $0 \le \varepsilon \ll 1$. In this regime, if the decomposition basis is matched with the system PSF, i.e. $w_0^2=\sigma^2$, the FI of the SpaDe measurement $\mathcal{I}_{\mathrm{SpaDe}}\left(\theta^2\right) = \mathcal{K}\left(\theta^2\right) [1-\varepsilon/2 + O(\varepsilon^2)]$ attains the QFI in the limit of $\varepsilon \to 0$. The PSF-matched binary SpaDe measurement outperforms the DI in the weak-scintillation regime with $\theta^2\lesssim 1.84 \, \sigma^2$; see Fig.~\ref{Fig:Gaussian_theta}. 

\begin{figure}[ht!]
    \centering
    \includegraphics[width=\linewidth]{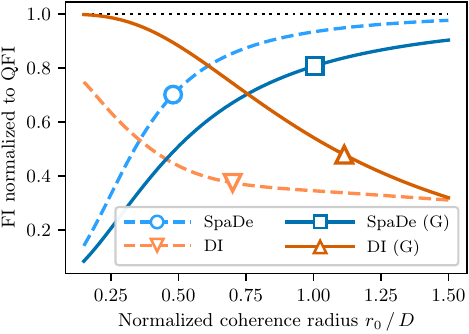}
    \caption{\label{Fig:Gaussian_r0} Ratio of the FI to the QFI, $\mathcal{I}(r_0)/\mathcal{K}(r_0)$, as a function of the spatial coherence radius $r_0$, for DI and binary SpaDe. 
    Solid lines correspond to the Gaussian (G) approximation, while dashed lines show results obtained from the realistic model, cf. Eq.~(\ref{Eq:Gamma2_im}). }
\end{figure}

Since the FI and QFI for the parameters $\theta^2$ and $r_0$ are related via
    \begin{align}
        \mathcal{K}(r_0)= \left(\frac{\partial \theta^2}{\partial r_0}\right)^{\!2}\mathcal{K}\left(\theta^2\right),
        \quad
        \mathcal{I}(r_0)= \left(\frac{\partial \theta^2}{\partial r_0}\right)^{\!2}\mathcal{I}\left(\theta^2\right),
    \end{align}
we can employ Eq.~(\ref{Eq:curve1}) to evaluate these quantities.
The resulting FI and QFI as functions of $r_0$ are shown in Fig.~\ref{Fig:Gaussian_r0} (solid lines).
The results again demonstrate the inefficiency of DI in the weak-scintillation regime and the superior performance of the binary SpaDe for $r_0 \gtrsim 0.76D$.
The binary SpaDe exhibits a clear advantage when the spatial coherence radius exceeds the receiver aperture diameter.

\section{Fisher information for hard-aperture model}
\label{Sec:FisherInfHard}
Let us consider the exact form of the density operator $\hat{\rho}_1$, defined by the second-order field correlation function $\Gamma_{2}^{(\mathrm{im})}(\mathbf{r}_1,\mathbf{r}_2)$; cf. Eq.~(\ref{Eq:Gamma2_im}).
In this case, the calculations are performed numerically.
In particular, the evaluation of the QFI relies on the geometric structure of quantum state space \cite{hubner1992,Braunstein1994,braunstein1996,pinel2013,marian2016}, where the QFI is directly related to the Bures distance \cite{Bures1969}.
Operationally, this amounts to computing the quantum fidelity between two infinitesimally close states, $\hat{\rho}_1(r_0)$ and $\hat{\rho}_1(r_0+\mathrm{d}r_0)$.  
The numerical procedure follows that described in Appendix~\ref{App:B} for the Gaussian approximation.

In the realistic case, the PSF is given by the Airy mode, 
\begin{align}\label{Eq:Airy}
     u^{\mathrm{(Airy)}}(\mathbf{r})
     = \frac{1}{\sqrt{\pi}} \frac{\BesselJ_1\! \left(\frac{\pi D}{\lambda f} r\right)}{r}, 
    \end{align}
where $\BesselJ_{m}\! \left(x\right)$ is the Bessel function of the first kind of order $m$ and $r=|\mathbf{r}|$.
We therefore consider binary SpaDe based on projection onto this mode (corresponding to the measurement outcome $n=0$) and its complementary mode space ($n=1$).
The probability that a photon passing through the aperture is projected onto the Airy mode is given by
    \begin{align}\label{Eq:pr0}
        P(n=0|r_0)&=\frac{S}{S_{\mathrm{ap}}^2}\int_{\mathcal{A}}\dd^2 \mathbf{r}_1\int_{\mathcal{A}}\dd^2 \mathbf{r}_2\, \Gamma_2^{\textrm{(ap)}}(\mathbf{r}_1, \mathbf{r}_2)\\
        &=\frac{16}{\pi} \int\limits_0^1  t \, \dd t \,
        \exp\!\left[ -\left(\frac{t D}{r_0}\right)^{5 / 3} \right] \nonumber\\  
        &\times\left( \frac{\pi}{2} - t \sqrt{1-t^2} - \arctan \! \left[ \frac{t}{\sqrt{1-t^2}} \right] \right).\nonumber
    \end{align}
The FI of the binary process with probability $P(0|r_0)$ can be readily evaluated and converted into the Fisher information for the coherence radius $r_0$.

The corresponding results are shown in Fig.~\ref{Fig:Gaussian_r0} (dashed lines).
Qualitative behavior of the QFI and FI for DI and binary SpaDe closely resembles that obtained with the Gaussian approximation: the DI becomes ineffective in the weak-scintillation regime, whereas binary SpaDe exhibits a clear performance advantage. 
Quantitatively, SpaDe detection is even more efficient in the realistic model, outperforming the DI for $r_0 \gtrsim 0.37D$.

\section{Numerical simulations}
\label{Sec:Simmulations}

To demonstrate the practical applicability of our method, we perform numerical simulations of a proposed experiment.
Light propagation through a turbulent atmosphere is modeled using the sparse-spectrum model \cite{Charnotskii2013a,Charnotskii2013b,Charnotskii2020} of the phase-screen method \cite{Fleck1976,Frehlich2000,Lukin_book,Schmidt_book,klen2023}.
Random temporal changes of the field amplitude is simulated using Taylor's frozen-turbulence hypothesis \cite{taylor1938} as consecutive wind-driven shifts of phase screens~\cite{Klen2024}.

We generate $N$ realizations of the unnormalized field amplitude with boundary condition $\widetilde{u}_i(\mathbf{r})=1$ at the emitting plane, corresponding to consecutive time steps enumerated by the index $i=1\ldots N$.
Propagation through the atmosphere yields aperture-plane field amplitudes $\widetilde{u}_i^{(\mathrm{ap})}(\mathbf{r})\equiv u_i^{(\mathrm{ap})}(\mathbf{r})\sqrt{S}$.
For each realization, we compute the probability of photon arrival, transmission through the aperture, and detection by one of the two detectors with the detection efficiency $\eta_{\mathrm{c}}$,
	\begin{align}\label{Eq:ProbW}
		W_i=\nu\int_{\mathcal{A}}\dd^2\mathbf{r}\, |\widetilde{u}_i^{\mathrm{(ap)}}(\mathbf{r})|^2.
	\end{align}
where $\nu=\eta_{\mathrm{c}} n_{\mathrm{th}}/S$ is the surface density of detected photons, as well as the conditional probability of detection in the Airy mode,
	\begin{align}\label{Eq:ProbSim}
		P_i=\frac{1}{S_{\mathrm{ap}}\displaystyle\int_{\mathcal{A}}\dd^2\mathbf{r}\, |\widetilde{u}_i^{\mathrm{(ap)}}(\mathbf{r})|^2}\left|\int_{\mathcal{A}}\dd^2 \mathbf{r}\, \widetilde{u}_i^{\mathrm{(ap)}}(\mathbf{r}) \right|^2.
	\end{align}
For details, see Appendix~\ref{App:D}.

For each realization, a detection event is sampled with probability $W_i$. 
Conditioned on detection, the outcome $n_i$ is assigned as $n_i=0$ with probability $P_i$ (Airy mode) and $n_i=1$ otherwise. 
This yields a set of $M\ll N$ detection events ${n_i}$, which we renumber as $i=1,\ldots,M$. 
The time step is chosen such that successive detection events are approximately statistically independent.

\begin{figure}[ht!]
    \centering
    \includegraphics[width=\linewidth]{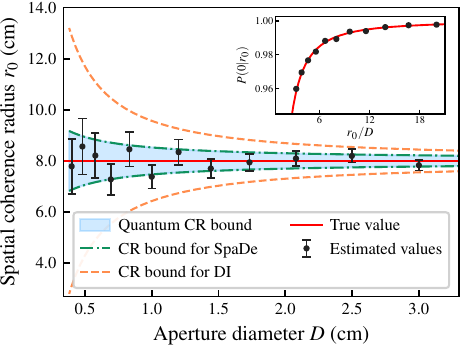}
    \caption{\label{Fig:Estimation} Results of the numerical simulation of light propagation and subsequent estimation of $r_0$ using binary detection.
    The variance asymptotically approaches the CR bound for the SpaDe.
    Parameters: propagation length $1~\textrm{km}$, refractive index structure constant $C_n^2=3.9\times 10^{-16}~\textrm{m}^{-2/3}$, wavelength $578~\textrm{nm}$, wind velocity $10~\mathrm{m/s}$, time step $0.2~\mathrm{ms}$, acquisition time $100~s$, mean photon number within the aperture $\nu S_{\mathrm{ap}}=0.02$, and sample size $M=8940$.
    Inset: $P(0|r_0)$ vs $r_0/D$, cf. Eq.~(\ref{Eq:pr0}), with sampled points. 
    }
\end{figure}

From the sample ${n_i}$, we estimate the binary-process parameter as $P \approx 1 - M^{-1}\sum_{i=1}^M n_i$, and convert it into an asymptotically unbiased estimator of $r_0$ using Eq.~(\ref{Eq:pr0}). 
This procedure is repeated for different aperture sizes. 
The results are shown in Fig.~\ref{Fig:Estimation}. 
To enable a fair comparison, we equalize the number of detected photons by fixing $M$ to its minimal value across apertures and discarding excess events. 
The binary SpaDe method clearly outperforms direct imaging for apertures much smaller than the coherence radius.

\section{Conclusions}
\label{Sec:Conclusions}

Estimation of the spatial coherence radius $r_0$ for light propagating through a turbulent atmosphere is a routine procedure when the diameter of the receiver aperture is much larger than $r_0$. 
In such cases, the estimation is usually performed by straightforward processing of data obtained via DI of a point-like source.
However, if the aperture diameter is smaller than the spatial coherence radius, diffraction effects significantly degrade the image of the source. 
As a result, estimation of $r_0$ becomes considerably more challenging. 

We analyze this situation within the framework of parameter estimation theory in the weak-field regime. 
The results show that, for small apertures, the FI associated with DI is far below the QFI, demonstrating that DI fails to extract much of the information about $r_0$ carried by the transmitted field.

The SpaDe detection technique, even in its simplest binary configuration, enables a significant improvement in the precision of $r_0$ estimation.
For the coherence radius significantly exceeding the input aperture, the corresponding FI saturates the QFI, rendering this procedure optimal.
We have verified our analytical predictions through numerical simulations based on a realistic modeling of the temporal random variations of the electromagnetic field in a turbulent atmosphere using the phase-screen method.
We believe that our results contribute to the further development of atmospheric turbulence estimation by introducing methods of quantum metrology into this field.
Moreover, they open a pathway toward superresolution imaging of sources separated from the observer by an atmospheric layer.

\section{Acknowledgment}
A.H. and M.K. acknowledge support from the National Research Foundation of Ukraine under Project No. 2023.03/0165.
A.A.S. acknowledges support from the National Academy of Sciences of Ukraine through Project No. 0125U000031, from Simons Foundation International SFI-PD-Ukraine-00014573, PI LB, and from the Virtual Ukraine Institute for Advanced Study (VUIAS) as a 2025/2026 Fellow. I.K. acknowledges support from the
Fondation de l’Ecole Polytechnique under contract XQUANT (2025-2027).

\appendix

\section{Turbulence models}
\label{App:A}
In the most general form, Eq.~(\ref{Eq:Gamma2_im_gen}) is given by (see  Refs.~\cite{Tatarskii2016,tatarskii1971,Fante1975,Fante1980,Andrews_book})
    \begin{align}\label{Eq:Gamma2_from}
        \Gamma_2^{\textrm{(ap)}}(\mathbf{r}_1, \, \mathbf{r}_2) = \frac{1}{S} \exp \left[ - \frac{1}{2} \mathcal{D}(\mathbf{r}) \right],
    \end{align}
where $ \mathbf{r} = \mathbf{r}_1 - \mathbf{r}_2 $ and $\mathcal{D}(\mathbf{r})$ is the wave structure function.
Assuming week scintillations and applying the Rytov approximation, the structure function for a spherical wave reads
 \begin{align}\label{Eq:WSF_sp}
    \mathcal{D}(\mathbf{r}) = 8\pi^2 k^2 \int\limits_{0}^{L}\dd z \int\limits_{0}^{\infty} \kappa \,\dd \kappa\, \Phi_n(\kappa;z)\left[1 - \BesselJ_0\left(\frac{\kappa z r}{L}\right)\right],
    \end{align}   
where $L$ is the propagation distance and $k=2 \pi/\lambda$ is the wavenumber.
The turbulence spectrum in the generalized von K\'{a}rm\'{a}n form is
    \begin{align}\label{Eq:Karman}
		\Phi_n(\kappa;z)=\frac{0.033\, C_n^2(z)\exp\left[-\left(\frac{\kappa\ell_0(z)}{5.92}\right)^2\right]}{\left\{\kappa^2+[2\pi/L_0(z)]^{2}\right\}^{11/6}}.
	\end{align}
Here $C_n^2(z)$ is the refractive-index structure constant, while $\ell_0(z)$ and $L_0(z)$ are inner and outer turbulence scales, respectively.

The Kolmogorov model of turbulence corresponds to $\ell_0(z)=0$ and $L_0(z)=+\infty$.
In this case, Eq.~(\ref{Eq:Gamma2_from}) reduces to Eq.~(\ref{Eq:Gamma2_im_gen}),
with
 \begin{align}\label{Eq:r0}
     r_0 = \left[1.46 \, k^2\int\limits_0^L\dd z\,  C_n^2(z)\left(\frac{z}{L}\right)^{5/3}\right]^{-3/5}.
 \end{align}
We may also assume that the source is located outside the atmosphere, such that $z/L\approx 1$ throughout the region where $C_n^2(z)\neq 0$.
This case corresponds to plane-wave propagation through the atmosphere.

In a more general scenario, one should account for finite values of $\ell_0$ and $L_0$---for example, when they are independent of $z$.
The estimation technique can be straightforwardly extended to this case.
One may still estimate the parameter $r_0$ defined by Eq.~(\ref{Eq:r0}), although, beyond the Kolmogorov model, its interpretation as the coherence radius is not generally rigorous.
If $\ell_0$ and $L_0$ are assumed to be known a priori, the problem remains single-parameter estimation.
More generally, one may consider simultaneous estimation of all three parameters.

\section{Interpolation formulas for Gaussian approximation}
\label{App:B}
Let us discuss the derivation of Eq.~(\ref{Eq:curve1}), describing the dependence of the parameters $\theta^2$ and $\sigma^2$ on the coherence radius $r_0$.
We start with calculating the quantum fidelity,
    \begin{align}\label{Eq:Fidelity}
        \mathcal{F}(\hat{\rho}_{1}, \hat{\rho}_{\textrm{(G)}}) =\left[\operatorname{Tr}\sqrt{\hat{\rho}_1^{1/2}\hat{\rho}_{\textrm{(G)}}\hat{\rho}_1^{1/2}}\right]^2,
    \end{align}
between the state $\hat{\rho}_1$ with position representation $\Gamma_2^{\textrm{(ap)}}(\mathbf{r}_1, \, \mathbf{r}_2)$ and its Gaussian approximation $\hat{\rho}_{\textrm{(G)}}$ with position representation $\Gamma_2^{\textrm{(G)}} (\mathbf{r}_1, \mathbf{r}_2)$.
To this end, we use the fact (see, e.g., Ref.~\cite{nielsen2010}) that the eigenvalues of $\sqrt{\hat{\rho}_1^{1/2}\hat{\rho}_{\mathrm{(G)}}\hat{\rho}_1^{1/2}}$ coincide with the singular values of $\sqrt{\hat{\rho}_1}\sqrt{\hat{\rho}_{\mathrm{(G)}}}$.
Knowledge of these eigenvalues therefore determines the quantum fidelity~(\ref{Eq:Fidelity}).
The numerical calculations are performed in the position representation on a grid chosen according to the Nyquist--Shannon theorem.

\begin{figure}[ht!]
    \centering
    \includegraphics[width=\linewidth]{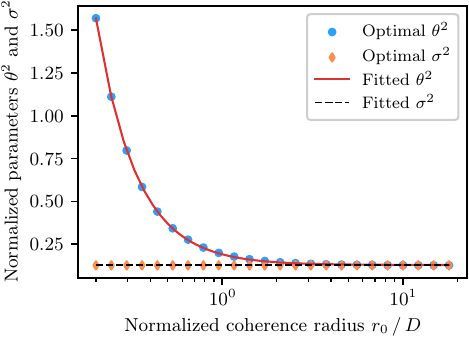}
    \caption{\label{Fig:Optimization} Parameters $\sigma^2$ and $\theta^2$, normalized to $(\lambda f/D)^2$, vs the normalized coherence radius $r_0/D$. The parameters are obtained by maximizing the quantum fidelity and subsequently fitted by Eq.~(\ref{Eq:curve1}). 
    }
\end{figure}

For the optimization procedure, we consider the density operator $\hat{\rho}_1$ [i.e., the correlation function $\Gamma_2^{\textrm{(ap)}}(\mathbf{r}_1,\mathbf{r}_2)$] for a given value of $r_0/D$, using units $\lambda f/D$ for $\mathbf{r}_{1,2}$.
We then determine the values of $\sigma^2$ and $\theta^2$ in the Gaussian density operator $\hat{\rho}_{\mathrm{(G)}}$ [i.e., the correlation function $\Gamma_2^{\textrm{(G)}}(\mathbf{r}_1,\mathbf{r}_2)$] that maximize the quantum fidelity~(\ref{Eq:Fidelity}).
As an initial guess, we use the values of $\theta^2$ and $\sigma^2$ corresponding to the absence of atmospheric turbulence, where the state is pure and $\theta^2=\sigma^2$.
In this case, the quantum fidelity can be calculated analytically,
    \begin{align}
        \mathcal{F}(\hat{\rho}_{1}, \hat{\rho}_{\textrm{(G)}})
        = \frac{2 \lambda^2 f^2}{\sigma^2 \pi^2 D^2} \left( 1 - \exp\left[- \frac{\sigma^2  \pi^2 D^2}{\lambda^2 f^2}\right] \right)^2.
    \end{align}
Its maximum is determined by a transcendental equation and is approximately attained at
    \begin{align}
        \sigma^2\approx\frac{2.5 \lambda^{2} f^{2}}{2 \pi^{2} D^{2}}.
    \end{align}
The resulting optimization and the fit by Eq.~(\ref{Eq:curve1}) are shown in Fig.~\ref{Fig:Optimization}.

\section{Derivation of Fisher information for the soft-aperture model}
\label{App:C}
The Gaussian approximation of the single-photon state, $\hat{\rho}_{\mathrm{(G)}}$, is characterized by the covariance matrix
    \begin{align}
        V=\mathrm{diag}\!\left(\theta^2,\,\frac{1}{4\sigma^2},\,\theta^2,\,\frac{1}{4\sigma^2}\right)
    \end{align}
for position and wave-vector $\begin{pmatrix} x&k_x&y&k_y \end{pmatrix}$. 
In this case, the QFI is expressed in terms of the quantum fidelity as (see Ref.~\cite{Bures1969}) 
    \begin{align}\label{Eq:QFI_G}
        \mathcal{K}(\theta^2) =
        -4\left.\frac{\dd^2}{\dd t^2}
        \sqrt{
        \mathcal{F}\!\left[
        \hat{\rho}(\theta^2),
        \hat{\rho}(\theta^2+t)
        \right]
        }
        \right|_{t=0}.
    \end{align}
For Gaussian states characterized by covariance matrices $V^{\prime}$ and $V^{\prime \prime}$, the fidelity reads \cite{marian2016}
    \begin{align}\label{Eq:Gaussian_fidelity}
        \mathcal{F}(\hat{\rho}^{\,\prime},\hat{\rho}^{\,\prime \prime})=\frac{2}{\left(\sqrt{K_{+}}-\sqrt{K_{-}}\right)^2},
    \end{align}
where $K_{\pm}=\sqrt{\Gamma}+\sqrt{\Lambda}\pm\sqrt{\Delta}$, and
    \begin{align}
        \Delta &= \det(V^{\prime}+V^{\prime \prime}),\\
        \Gamma &= 2^4\det\!\left((JV^{\prime})(JV^{\prime \prime})-\frac{1}{4}\mathbb{I}_4\right),\\
        \Lambda &= 2^4\det\!\left(V^{\prime}+\frac{i}{2}J\right)\det\!\left(V^{\prime \prime}+\frac{i}{2}J\right),
    \end{align}
and 
\begin{align}
J=\mathbb{I}_2\otimes
\begin{pmatrix}
0 & 1\\
-1 & 0
\end{pmatrix}.
\end{align}
Substituting Eq.~(\ref{Eq:Gaussian_fidelity}) into Eq.~(\ref{Eq:QFI_G}) yields Eq.~(\ref{Eq:QFI_gauss}).

The FI for the DI scenario is obtained straightforwardly, since it is equivalent to evaluating the FI for the variance of a centered Gaussian distribution.
For the SpaDe scenario, we consider a binary process, characterizing by the probability of a photon projection onto the Gaussian beam, 
    \begin{align}
        P(n=0| \theta^2) = \frac{4 \sigma^2 w_0^2}{(\theta^2 + w_0^2)(\sigma^2 + w_0^2)}.
    \end{align}
Converting the FI of this binary process to the FI for $\theta^2$, we arrive at Eq.~(\ref{Eq:SpaDe_Gauss}).
In the considered case of PSF-matched SpaDe $w_0^2 = \sigma^2$, this reduces to
     \begin{align}
        \mathcal{I}_{\mathrm{SpaDe}}(\theta^2) = \frac{2 \, \sigma^2 }{(\theta^4 - \sigma^4) \, (\theta^2 + \sigma^2)}.
     \end{align}
This FI approaches the quantum limit (\ref{Eq:QFI_gauss}) for weak scintillations.

\section{Numerical simulations details}
\label{App:D}
Here we derive Eqs.~(\ref{Eq:ProbW}) and (\ref{Eq:ProbSim}) describing, respectively, the photon-arrival probability after transmission through the aperture and detection by one of the two detectors, and the conditional probability of detection in the Airy mode.
To this end, we first rewrite the instantaneous quantum state (\ref{Eq:DensOp_Inst}) as
    \begin{align}\label{Eq:DensOp_Inst_A}
        \hat{\rho}_{\mathrm{inst}}&=(1-\eta n_{\mathrm{th}})\dyad{\mathrm{vac}}{\mathrm{vac}}+n_{\mathrm{th}}\,\eta\,\hat{\rho}_{\mathrm{inst};1},
    \end{align}
where
    \begin{align}\label{Eq:Single-Photon-Instant}
        \hat{\rho}_{\mathrm{inst};1}=
        \frac{1}{\eta}
        \int_{\mathbb{R}^4}\dd^2\mathbf{r}_1\dd^2\mathbf{r}_2\,
       u^{\textrm{(im)}}(\mathbf{r}_1)\, [u^{\textrm{(im)}}(\mathbf{r}_2)]^{\ast}\,\dyad{\mathbf{r}_1}{\mathbf{r}_2}
    \end{align}
is the instantaneous single-photon operator.

The factor $\eta n_{\mathrm{th}}$ can be interpreted as the probability $W$ in Eq.~(\ref{Eq:ProbW}). 
Substituting Eq.~(\ref{Eq:ImageField}) into Eq.~(\ref{Eq:eta}), we obtain
  \begin{align}
    \eta=\int_{\mathcal{A}}\dd^2\mathbf{r} |u^{\textrm{(ap)}}(\mathbf{r})|^2.  
  \end{align}  
Rewriting this expression in terms of $\widetilde{u}^{\textrm{(ap)}}(\mathbf{r})$ and accounting for the detection efficiency $\eta_{\mathrm{c}}$, we arrive at Eq.~(\ref{Eq:ProbW}).
Next, probability of projecting the instanteneous state onto the Airy mode is given by
    \begin{align}
        P=\bra{\mathrm{Airy}}\hat{\rho}_{\mathrm{inst};1}\ket{\mathrm{Airy}},
    \end{align}   
where
    \begin{align}
        \ket{\mathrm{Airy}}=\int_{\mathbb{R}^2}\dd^2 \mathbf{r}\, u^{\mathrm{(Airy)}}(\mathbf{r})\ket{\mathbf{r}}
    \end{align}    
and $u^{\mathrm{(Airy)}}(\mathbf{r})$ is defined by Eq.~(\ref{Eq:Airy}).
Utilizing here Eqs.~(\ref{Eq:Single-Photon-Instant}) and (\ref{Eq:ImageField}), and rewriting the obtained expression in terms of $\widetilde{u}^{\textrm{(ap)}}(\mathbf{r})$, we arrive at Eq.~(\ref{Eq:ProbSim}).

\bibliography{biblio}

\end{document}